\input{aipcheck}

\documentclass[
    ,final            
  ]
  {aipproc}

\layoutstyle{8x11single}

\begin{document}

\title{Simultaneous Description of Even-Even, Odd-Mass and Odd-Odd Nuclear Spectra}


\author{\underline{H. G. Ganev}}{
  address={Institute of Nuclear Research and Nuclear Energy,
Bulgarian Academy of Sciences, Sofia 1784, Bulgaria} }

\author{A. I. Georgieva}{
  address={Institute of Nuclear Research and Nuclear Energy,
Bulgarian Academy of Sciences, Sofia 1784, Bulgaria} }

\begin{abstract}
 The orthosymplectic extension of the Interacting Vector Boson Model
(IVBM) is used for the simultaneous description of the spectra of
different families of neighboring heavy nuclei. The structure of
even-even nuclei is used as a core on which the collective
excitations of the neighboring odd-mass and odd-odd nuclei are built
on. Hence, the spectra of the odd-mass and odd-odd nuclei arise as a
result of the consequent and self-consistent coupling of the fermion
degrees of freedom of the odd particles, specified by the fermion
sector $SO^{F}(2\Omega)\subset OSp(2\Omega/12,R)$, to the boson core
which states belong to an $Sp^{B}(12,R)$ irreducible representation.

The theoretical predictions for different low-lying collective bands
with positive and negative parity for two sets of neighboring nuclei
with distinct collective properties are compared with experiment and
IBM/IBFM/IBFFM predictions. The obtained results reveal the
applicability of the used dynamical symmetry of the model.

\end{abstract}

\maketitle


\section{Introduction}

In recent years, extensive experimental evidence for the existence
of distinct band structures in different even-even, odd-mass and
odd-odd nuclei has been obtained. It has created an opportunity for
testing the predictions of different theoretical models on the level
properties of these nuclei. Generally, it is believed that the
properties of the odd-mass nuclei can be described using
particle-core coupled-type models. In this respect, it was found in
nuclear physics that the combined algebras of bosons and fermions
are particular useful in a description of the spectra of the
odd-mass nuclei. Moreover, an attempt to get a classification of
several nuclei including even and odd ones have been realized and
very interesting results have been obtained \cite{IBM}, \cite{IBFM}.
The main attention has been paid to the investigations of the models
with the Hamiltonians, which exhibit dynamical supersymmetries. Such
Hamiltonians are presented by the sums of the Casimir operators for
an appropriated subgroup chain of a certain supergroup. The spectra
of even-even and odd-mass nuclei, which states are unified into a
common supermultiplet, are described by the same Hamiltonian with a
single set of adjustable parameters.

In the early 1980s, a boson-number-preserving version of the
phenomenological algebraic Interacting Vector Boson Model (IVBM)
\cite{IVBM} was introduced and applied successfully \cite{IVBMrl} to
a description of the low-lying collective rotational spectra of the
even-even medium and heavy mass nuclei. With the aim of extending
these applications to incorporate new experimental data on states
with higher spins and to incorporate new excited bands, we explored
the symplectic extension of the IVBM \cite{GGG}, for which the
dynamical symmetry group is $Sp^{B}(12,R).$ This extension is
realized from, and has its physical interpretation over basis states
of its maximal compact subgroup $U^{B}(6)\subset Sp^{B}(12,R)$, and
resulted in the description of various excited bands of both
positive and negative parity of complex systems exhibiting
rotation-vibrational spectra \cite{Spa}. In \cite{OSE} an
orthosymplectic extension of the IVBM was carried out in order to
encompass the treatment of the odd-mass nuclei. In \cite{DB} the
investigation of the doublet bands in doubly odd nuclei from
$A\sim130$ region was presented. A good agreement between experiment
and theoretical predictions for the energy levels of these bands as
well as in-band $B(E2)$ and $B(M1)$ transition probabilities is
obtained. With the present work we exploit further the boson-fermion
extension \cite{OSE} of IVBM for the simultaneous description of the
low-lying positive and/or negative bands of two different sets of
neighboring even-even, odd-mass and odd-odd nuclei with various
collective properties. The first set of nuclei which we choose
possesses a clear rotational character, whereas the second one
displays typical vibrational features. The theoretical description
of the doubly odd nuclei under consideration is fully consistent and
starts with the calculation of theirs even-even and odd-mass
neighbors. We consider the simplest physical picture in which two
particles (or quasiparticles) with intrinsic spins taking a single
$j-$value are coupled to an even-even core nucleus which states
belong to an $Sp^{B}(12,R)$ irreducible representation (irrep).
Thus, the bands of the odd-mass and odd-odd nuclei arise as
collective bands built on a given even-even nucleus. Nevertheless,
the results for the energy spectra obtained in this simplified
version of the model agree rather well with the experimental data.

\section{The even-even core nuclei}

The algebraic structure of the IVBM is realized in terms of creation
and annihilation operators of two kinds of vector bosons
$u_{m}^{+}(\alpha )$, $u_{m}(\alpha )$ ($m=0,\pm 1$), which differ
in an additional quantum number $\alpha=\pm1/2-$the projection of
the $T-$spin (an analogue to the $F-$spin of IBM-2). The bilinear
products of the creation and annihilation operators of the two
vector bosons generate the boson representations of the non-compact
symplectic group $ Sp^{B}(12,R)$ \cite{IVBM}:
\begin{eqnarray}
F_{M}^{L}(\alpha ,\beta ) &=& {\sum }_{k,m}C_{1k1m}^{LM}u_{k}^{+}(
\alpha )u_{m}^{+}(\beta ),  \nonumber \\
G_{M}^{L}(\alpha ,\beta ) &=&{\sum }_{k,m}C_{1k1m}^{LM}u_{k}(\alpha
)u_{m}(\beta ),  \label{pairgen}
\end{eqnarray}
\begin{equation}
A_{M}^{L}(\alpha, \beta )={\sum }_{k,m}C_{1k1m}^{LM}u_{k}^{+}(\alpha
)u_{m}(\beta ),  \label{numgen}
\end{equation}
where $C_{1k1m}^{LM}$, which are the usual Clebsch-Gordon
coefficients for $L=0,1,2$ and $M=-L,-L+1,...L$, define the
transformation properties of (\ref{pairgen}) and (\ref{numgen})
under rotations. Being a noncompact group, the unitary
representations of $Sp^{B}(12,R)$ are of infinite dimension, which
makes it impossible to diagonalize the most general Hamiltonian.
When restricted to the group $U^{B}(6)$, each irrep of the group
$Sp^{B}(12,R)$ decomposes into irreps of the subgroup characterized
by the partitions \cite{GGG},\cite{Q1}: $\lbrack
N,{0}^5\rbrack_{6}\equiv \lbrack N]_{6}$, where $N=0,2,4,\ldots$
(even irrep) or $N=1,3,5,\ldots$ (odd irrep). The subspaces
$[N]_{6}$ are  finite dimensional, which simplifies the problem of
diagonalization. Therefore the complete spectrum of the system can
be calculated through the diagonalization of the Hamiltonian in the
subspaces of all the unitary irreducible representations (UIR) of
$U^{B}(6)$, belonging to a given UIR of $Sp^{B}(12,R)$, which
further clarifies its role of a group of dynamical symmetry.

The most important application of the $U^{B}(6)\subset $
$Sp^{B}(12,R)$ limit of the theory is the possibility it affords for
describing both even and odd parity bands up to very high angular
momentum \cite{GGG}. In order to do this we first have to identify
the experimentally observed bands with the sequences of basis states
of the even $Sp^{B}(12,R)$ irrep (Table \ref{BasTab}). As we deal
with the symplectic extension we are able to consider all even
eigenvalues of the number of vector bosons $N$ with the
corresponding set of $T-$spins, which uniquely define the
$SU^{B}(3)$ irreps $(\lambda, \mu)$. The multiplicity index $K$
appearing in the final reduction to the $SO^{B}(3)$ is related to
the projection of $L$ on the body fixed frame and is used with the
parity ($\pi $)\ to label the different bands ($K^{\pi } $) in the
energy spectra of the nuclei. For the even-even nuclei we have
defined the parity of the states as $\pi_{core} =(-1)^{T}$
\cite{GGG}. This allowed us to describe both positive and negative
bands.

Further, we use the algebraic concept of \textquotedblleft
yrast\textquotedblright\ states, introduced in \cite{GGG}. According
to this concept we consider as yrast states the states with given
$L$, which minimize the energy with respect to the number of vector
bosons $N$ that build them. Thus the states of the ground state band
(GSB) were identified with the $SU^{B}(3)$ multiplets $(0,\mu )$
\cite{GGG}. In terms of $(N,T)$ this choice corresponds to $(N=2\mu
,T=0)$ and the sequence of states with different numbers of bosons
$N=0,4,8,\ldots $ and $T=0$ (and also $T_{0}=0$). The presented
mapping of the experimental states onto the $SU^{B}(3)$ basis
states, using the algebraic notion of yrast states, is a particular
case of the so called "stretched" states \cite{str}. The latter are
defined as the states with ($\lambda_{0}+2k,\mu_{0}$) or
($\lambda_{0},\mu_{0}+ k$), where $N_{i}=\lambda_{0}+2\mu_{0}$ and
$k=0,1,2,3, \ldots$. In the symplectic extension of the IVBM the
change of the number $k$, which is related in the applications to
the angular momentum $L$ of the states, gives rise to the collective
bands.

It was established \cite{GGD} that the correct placement of the
bands in the spectrum strongly depends on their bandheads'
configuration, and in particular, on the minimal or initial number
of bosons, $N = N_{i}$, from which they are built. The latter
determines the starting position of each excited band. In the
present application we take for $N_{i}$ the value at which the best
$\chi^2$ is obtained in the fitting procedure for the energies of
the considered excited band.

\begin{figure}[h]
\includegraphics[height=0.56\textheight]{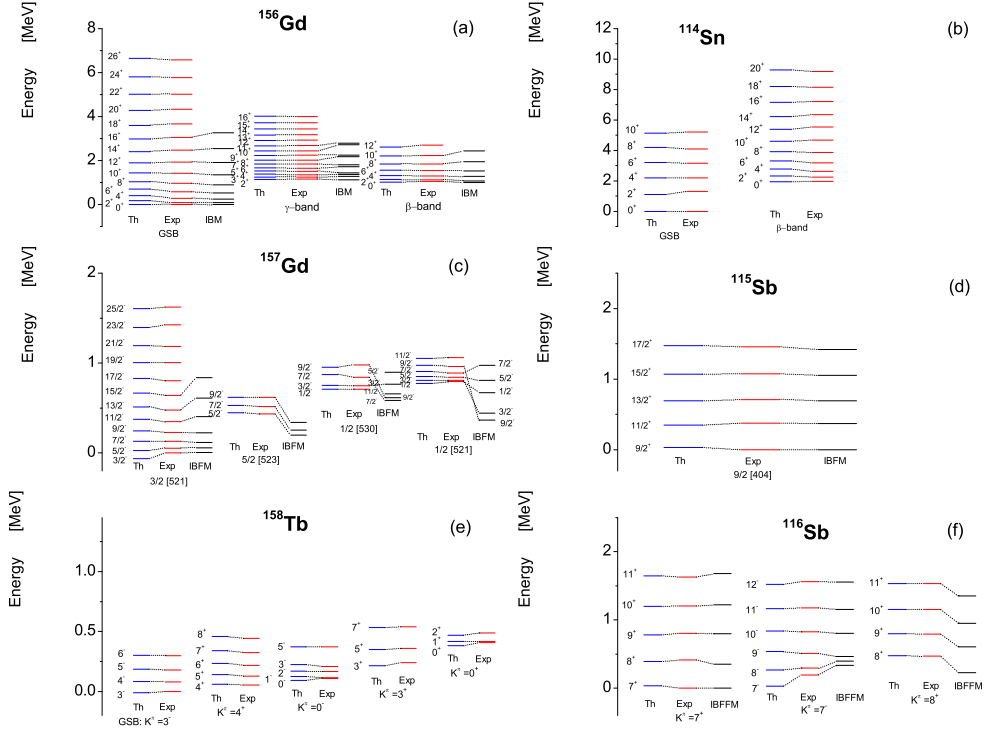}
\caption{Comparison of the theoretical, experimental and
IBM/IBFM/IBFFM energies for the ground and first excited positive
and/or negative parity bands of $^{156}$Gd (a),$^{157}$Gd (c),
$^{158}$Tb (e), ground and first excited $\beta$ bands of $^{114}$Sn
(b), $K^{\pi}_{J}=9/2^{+}$ intruder band of $^{115}$Sb (d), and the
intruder $K^{\pi}_{J}=7^{+}$, $K^{\pi}_{J}=7^{-}$ and
$K^{\pi}_{J}=8^{+}$ bands of $^{116}$Sb (f).}\label{all}
\end{figure}

To describe the structure of odd-mass and odd-odd nuclei, first a
description of the appropriate even-even cores should be obtained.
We use the unitary limit of IVBM \cite{GGG} and determine the values
of five phenomenological model parameters, related to the boson
degrees of freedom (see Eq. (\ref{Energy})), by fitting the energies
of the ground and few excited bands ($\gamma-$ and/or $\beta-$
bands) of the even-even nuclei to the experimental data \cite{exp},
using a $\chi^{2}$ procedure. The theoretical predictions for the
even-even core nuclei are presented in Figs. \ref{all}(a) and
\ref{all}(b). For comparison, the predictions of IBM are also shown.
From the figures one can see that the calculated energy levels agree
rather well up to very high angular momenta with the observed data.
One can see also that for high spins ($L\geq 10-14$), where the
deviations of the IBM predictions become more significant, the
structure of the energy levels of the GSB ($\beta-$ and
$\gamma-$bands) is reproduced rather well. The $^{156}Gd$ core
nucleus shows a well-developed ground state rotational band, whereas
the $^{114}Sn$ one shows a purely vibrational (equidistant) spectrum
as can be clearly seen from Fig. \ref{all}(b). The latter is
described in IBM in its $U(5)$ dynamical symmetry limit which is
reasonable for spherical nuclei in the $Sn-Sb$ region.


\section{Orthosymplectic extension}

In order to incorporate the intrinsic spin degrees of freedom into
the symplectic IVBM, we extend the dynamical algebra of
$Sp^{B}(12,R)$ to the orthosymplectic algebra of $OSp(2\Omega/12,R)$
\cite{OSE}. For this purpose we introduce a particle (quasiparticle)
with spin $j$ and consider a simple core plus particle picture.
Thus, in addition to the boson collective degrees of freedom we
introduce creation and annihilation operators $a_{m}^{\dag}$ and
$a_{m}$ ($m=-j,\ldots,j$), which satisfy the anticommutation
relations: $\{a_{m},a_{m'}^{\dag}\}=\delta _{mm'}$ and
$\{a_{m}^{\dag},a_{m'}^{\dag}\}=\{a_{m},a_{m'}\}=0$.

All bilinear combinations of \ $a_{m}^{+}$ and $a_{m'}$, namely
\begin{eqnarray}
f_{mm'} &=&a_{m}^{\dag}a_{m'}^{\dag}, \ \ m\neq m' \nonumber \\
g_{mm'} &=&a_{m}a_{m'}, \ \ m\neq m'; \label{O2N gen} \\
C_{mm'} &=&(a_{m}^{\dag}a_{m'}-a_{m'}a_{m}^{\dag})/2  \label{UN gen}
\end{eqnarray}%
generate the (Lie) fermion pair algebra of $SO^{F}(2\Omega)$. Their
commutation relations are given in \cite{OSE}. The number preserving
operators (\ref{UN gen}) generate maximal compact subalgebra of
$SO^{F}(2\Omega)$, i.e. $U^{F}(\Omega)$.

\subsection{Fermion dynamical symmetries}

One can further construct a certain fermion dynamical symmetry, i.e.
the group-subgroup chain:
\begin{equation}
SO^{F}(2\Omega) \supset  G ' \supset  G '' \supset \ldots
.\label{FDS}
\end{equation}
In particular for one particle occupying a single level $j$ we are
interested in the following dynamical symmetry:
\begin{equation}
SO^{F}(2\Omega) \supset  Sp(2j+1) \supset SU^{F}(2),\label{opFDS}
\end{equation}
where $Sp(2j+1)$ is the compact symplectic group. The dynamical
symmetry (\ref{opFDS}) remains valid also for the case of two
particles occupying the same level $j$. In this case, the allowed
values of the quantum number $I$ of $SU^{F}(2)$ in (\ref{opFDS})
according to reduction rules are $I=0,2,\ldots,2j-1$ \cite{pair}. If
the two particles occupy different levels $j_{1}$ and $j_{2}$ of the
same or different major shell(s), one can consider the chain
\begin{equation}
SO^{F}(2\Omega )\supset U^{F}(\Omega )%
\begin{tabular}{llllllll}
$\nearrow $ & $U^{F}(\Omega _{1})$ & $\supset $ & $Sp(2j_{1}+1)$ & $\supset $ & $%
SU^{F}_{I_{1}}(2)$ & $\searrow $ &  \\
&  &  &  &  &  &  & $SU^{F}(2)$ \\
$\searrow $ & $U^{F}(\Omega _{2})$ & $\supset $ & $Sp(2j_{2}+1)$ & $\supset $ & $%
SU^{F}_{I_{2}}(2)$ & $\nearrow $ &
\end{tabular}
\label{tpFDS}
\end{equation}
where $\Omega=\Omega_{1} + \Omega_{2}$. For simplicity hereafter we
will use just the reduction $SO^{F}(2\Omega) \supset SU^{F}(2)$
(i.e. dropping all intermediate subgroups between $SO^{F}(2\Omega)$
and $SU^{F}(2)$) and keep in mind the proper content of the set of
$I$ values for one and/or two particles cases, respectively.

\subsection{Dynamical Bose-Fermi symmetry. Dynamical supersymmetry}

Once the fermion dynamical symmetry is determined we proceed with
the construction of the Bose-Fermi symmetries. If a fermion is
coupled to a boson system having itself a dynamical symmetry (e.g.,
such as an IBM core), the full symmetry of the combined system is
$G^{B} \otimes G^{F}$. Bose-Fermi symmetries occur if at some point
the same group appears in both chains $G^{B} \otimes G^{F} \supset
G^{BF}$, i.e. the two subgroup chains merge into one.

The standard approach to supersymmetry in nuclei (dynamical
supersymmetry) is to embed the Bose-Fermi subgroup chain of $G^{B}
\otimes G^{F}$ into a larger supergroup G, i.e. $G \supset G^{B}
\otimes G^{F}$. Making use of the embedding $SU^{F}(2)\subset
SO^{F}(2\Omega)$ we make orthosymplectic (supersymmetric) extension
of the IVBM which is defined through the chain \cite{OSE}:
\begin{equation}
\begin{tabular}{lllll}
$OSp(2\Omega/12,R)$ & $\supset $ & $SO^{F}(2\Omega)$ & $\otimes $ & $Sp^{B}(12,R)$ \\
&  &  &  & $\ \ \ \ \ \ \Downarrow $ \\
&  & $\ \ \ \ \ \Downarrow $ & $\otimes $ & \ $\ U^{B}(6)$ \\
&  &  &  & $\ \ \ \ \ N$ \\
&  &  &  & $\ \ \ \ \ \ \Downarrow $ \\
&  & $SU^{F}(2)$ & $\otimes $ & $SU^{B}(3)\otimes U_{T}^{B}(2)$ \\
&  & $\ \ \ \ \ I$ &  & $(\lambda ,\mu )\Longleftrightarrow (N,T)~$ \\
&  & \multicolumn{1}{r}{$\searrow $} &  & $\ \ \ \ \ \ \Downarrow $ \\
&  &  & $\otimes $ & $SO^{B}(3)\otimes U^{B}(1)$ \\
&  &  &  & $~~~L\qquad \qquad T_{0}$ \\
&  &  & $\Downarrow $ &  \\
&  & $Spin^{BF}(3)$ & $\supset $ & $Spin^{BF}(2),$ \\
&  & $~~~~~J$ &  & $~~~~~J_{0}$%
\end{tabular}
\label{chain}
\end{equation}
where below the different subgroups the quantum numbers
characterizing their irreducible representations are given.
$Spin^{BF}(n)$ ($n=2,3$) denotes the universal covering group of
$SO(n)$.

\section{The energy spectra of odd-mass and odd-odd nuclei}

We can label the basis states according to the chain (\ref{chain})
as:
\begin{eqnarray}
|~[N]_{6};(\lambda,\mu);KL;I;JJ_{0};T_{0}~\rangle \equiv
|~[N]_{6};(N,T);KL;I;JJ_{0};T_{0}~\rangle ,  \label{Basis}
\end{eqnarray}
where $[N]_{6}-$the $U^{B}(6)$ labeling quantum number and
$(\lambda,\mu)-$the $ SU^{B}(3)$ quantum numbers characterize the
core excitations, $K$ is the multiplicity index in the reduction
$SU^{B}(3)\supset SO^{B}(3)$, $L$ is the core angular momentum,
$I-$the intrinsic spin of an odd particle (or the common spin of two
fermion particles for the case of odd-odd nuclei), $J,J_{0}$ are the
total (coupled boson-fermion) angular momentum and its third
projection, and $T$,$T_{0}$ are the $T-$spin and its third
projection, respectively.

The Hamiltonian of the combined boson-fermion system can be written
as linear combination of the Casimir operators of the different
subgroups in (\ref{chain}): \
\begin{eqnarray}
H = aN+bN^{2}+\alpha _{3}T^{2}+\beta _{3}^{\prime
}L^{2}+\alpha_{1}T_{0}^{2} + \eta I^{2}+\gamma ^{\prime }J^{2}+\zeta
J_{0}^{2} \label{Hamiltonian}
\end{eqnarray}
and it is obviously diagonal in the basis (\ref{Basis}) labeled by
the quantum numbers of their representations. Then the eigenvalues
of the Hamiltonian (\ref{Hamiltonian}), that yield the spectrum of
the odd-mass and odd-odd systems are:
\begin{eqnarray}
E(N;T,T_{0};L,I;J,J_{0})=aN+bN^{2}+\alpha _{3}T(T+1)+ \beta
_{3}^{\prime }L(L+1)+\alpha _{1}T_{0}^{2} +\eta I(I+1)+\gamma
^{\prime }J(J+1)+\zeta J_{0}^{2}.\label{Energy}
\end{eqnarray}
The infinite set of basis states classified according to the
reduction chain (\ref{chain}) are schematically shown in Table
\ref{BasTab}.
\begin{table}[h]
\caption{Classification scheme of basis states (\protect\ref{Basis})
according the decompositions given by the chain
(\protect\ref{chain}).}
\smallskip \centering{\footnotesize \renewcommand{\arraystretch}{1.25}
\begin{tabular}{l|l|l|l|l|l}
\hline\hline $N$ & $T$ & $(\lambda ,\mu )$ & $K$ & $L$ & \quad \quad
\quad \quad \quad \quad \quad $J=L\pm I$ \\ \hline\hline $0$ & $0$ &
$(0,0)$ & $0$ & $0$ & $I$ \\ \hline\hline $2$ & $1$ & $(2,0)$ & $0$
& $0,2$ & $I;\ 2\pm I$ \\ \cline{2-6} & $0$ & $(0,1)$ & $0$ & $1$ &
$1\pm I$ \\ \hline\hline & $2$ & $(4,0)$ & $0$ & $0,2,4$ & $I;\ 2\pm
I;4\pm I$ \\ \cline{2-6}
$4$ & $1$ & $(2,1)$ & $1$ & $1,2,3$ & $1\pm I;\ 2\pm I;\ 3\pm I$ \\
\cline{2-6} & $0$ & $(0,2)$ & $0$ & $0,2$ & $I;\ 2\pm I$ \\
\hline\hline
$\vdots $ & $\vdots $ & $\vdots $ & $\vdots $ & $\vdots $ & $\vdots $%
\end{tabular}
} \label{BasTab}
\end{table}

The basis states (\ref{Basis}) can be considered as a result of the
coupling of the orbital $\mid (N,T);KLM;T_{0}\rangle$ and spin
$\phi_{j\equiv I,m}$ wave functions. Then, if the parity of the
single particle is $\pi_{sp}$, the parity of the collective states
of the odd$-A$ nuclei will be $\pi=\pi_{core} \pi_{sp}$ \cite{OSE}.
In analogy, one can write $\pi=\pi_{core} \pi_{sp}(1) \pi_{sp}(2)$
for the case of odd-odd nuclei. Thus, the description of the
positive and/or negative parity bands requires only the proper
choice of the core band heads, on which the corresponding single
particle(s) is (are) coupled to, generating in this way the
different odd$-A$ (odd-odd) collective bands.

In our application, the most important point is the identification
of the experimentally observed states with a certain subset of basis
states from (ortho)symplectic extension of the model. In general,
except for the excited bands of the even-even nuclei for which the
stretched states of the first type ($\lambda-$changing) are used,
the stretched states of the second type ($\mu-$changing) are
considered in all the calculations of the collective states of the
odd-mass and doubly odd nuclei.

Additionally to the five parameters $a, b, \alpha_{3}, \beta_{3},
\alpha_{1}$ entering in Eq. (\ref{Energy}) which are used to
describe the energies of the even-even core nuclei, the number of
adjustable parameters needed for the complete description of the
collective spectra of both odd-A and odd-odd nuclei is three, namely
$\gamma$, $\zeta $ and $\eta$. The first two are evaluated by a fit
to the experimental data \cite{exp} of the GSB of the corresponding
odd-A neighbor, while the last one is introduced in the final step
of the fitting procedure for the odd-odd nucleus, respectively.

The odd-A nucleus $^{157}Gd$ can be considered as a neutron particle
coupled to the $^{156}Gd$ core, while the $^{115}Sb$ is obtained by
proton coupling to the $^{116}Sn$, respectively. The comparison
between the experimental spectra for the low-lying bands and our
calculations for $^{157}Gd$ and $^{115}Sb$ nuclei is illustrated in
Figs. \ref{all}(c) and \ref{all}(d). In our considerations we take
into account only the first available single particle orbit $j_{1}$.
One can see from the figures that the calculated energy levels agree
rather well in general with the experimental data up to very high
angular momenta. For comparison, in the figures the IBFM results are
also shown. Note that all calculated levels, for the bands
considered, are in correct order in contrast to IBFM results (for
$^{157}Gd$). Another difference between the IVBM and IBFM
predictions is that in the former the correct placement of all the
band heads is reproduced quite well.

For the calculation of the odd-odd nuclear spectra a second particle
should be coupled to the core. In our approach a consistent
procedure is employed which includes the analysis of the even-even
and odd-mass neighbors of the nucleus under consideration. Thus, as
a first step the first odd particle was coupled to the boson core in
order to obtain the spectra of the odd-mass neighbors $^{157}Gd$ and
$^{115}Sb$. As a second step, we consider the addition of a second
particle on a single $j_{2}$ level to the boson-fermion system. The
theoretical predictions for the ground and first excited bands of
$^{158}Tb$ and intruder bands of $^{116}Sb$ odd-odd nuclei are
presented in Figs. \ref{all}(e) and \ref{all}(f). They are compared
with the experimental data and for the $^{116}Sb$ case also with
IBFFM predictions. From the figures one can see the good overall
agreement between the theory and experiment which reveals the
applicability of the used orthosymplectic extension of the model.

\section{Conclusion}

In the present paper, the structure of different collective bands
with both positive and negative parity for two sets of neighboring
even-even, odd-mass and odd-odd nuclei with various collective
properties is presented. The even-even nuclei are used as a core on
which the collective excitations of the neighboring odd-mass and
odd-odd nuclei are built on. Thus, the spectra of odd-mass and
odd-odd nuclei arise as a result of the consequent and
self-consistent coupling of the fermion degrees of freedom to the
boson core. Hence, a purely collective structure of the bands under
consideration is introduced. The good agreement between the
theoretical and the experimental band structures confirms the
applicability of the used dynamical symmetry of the IVBM. The
success is based on the (ortho)symplectic structures of the model
which allow the mixing of the basic collective modes $-$rotational
and vibrational ones arising from the yrast conditions$-$ way back
on the level of the even-even cores. This allows for the correct
reproduction of the high spin states of the collective bands and the
correct placement of the different band heads. The simplifications
in our approach comes from the fact that only one dynamical symmetry
is employed, which leads to exact and simple solutions depending
only on the values of the model parameters. The success of the
presented application is based also mainly on the proper and
consistent mapping of the experimentally observed collective states
of the even-even, odd-mass and odd-odd nuclei on the
(ortho)symplectic structures. The latter is much simpler approach
than the mixing of the basis states considered in other theoretical
models.


\begin{theacknowledgments}
This work was supported by the Bulgarian National Foundation for
scientific research under Grant Number $\Phi -1501$. H. G. G.
acknowledges also the support from the European Operational programm
HRD through contract $BG051PO001/07/3.3-02$ with the Bulgarian
Ministry of Education.
\end{theacknowledgments}



\bibliographystyle{aipproc}   

\bibliography{sample}

\IfFileExists{\jobname.bbl}{}
 {\typeout{}
  \typeout{******************************************}
  \typeout{** Please run "bibtex \jobname" to optain}
  \typeout{** the bibliography and then re-run LaTeX}
  \typeout{** twice to fix the references!}
  \typeout{******************************************}
  \typeout{}
 }



\end{document}